\documentclass[10pt,aps,prb,superscriptaddress,twocolumn,a4paper,floatfix]{revtex4-2}
\usepackage{amsmath}%
\usepackage{amsfonts}%
\usepackage{amssymb}%
\usepackage[latin1]{inputenc}
\usepackage{graphicx}
\usepackage{color}
\usepackage[english]{babel}
\usepackage{natbib}
\usepackage[colorlinks=true, citecolor=blue, linkcolor=blue, urlcolor=blue]{hyperref}
\usepackage{booktabs}
\usepackage{enumitem}
\usepackage{subfigure}
\usepackage{multirow}
\usepackage{braket}
\usepackage{dcolumn}
\usepackage{ulem}
\usepackage{bm}
\usepackage[parse-numbers=true]{siunitx}
\usepackage{textgreek}
\usepackage{wasysym}

\definecolor{Ora}{cmyk}{0,0.6,0.8,0.4}

\graphicspath{{figures/}}
\usepackage{graphicx} 
\usepackage{float} 
\usepackage{xcolor}
\usepackage{gensymb} 

\begin{document}

\title{Intermixing-driven surface and bulk ferromagnetism in the quantum anomalous Hall candidate MnBi$_6$Te$_{10}$}

\author{Abdul~V.~Tcakaev}
\affiliation{Physikalisches Institut (EP-IV), Fakult\"{a}t f\"{u}r Physik und Astronomie, Universit\"{a}t W\"{u}rzburg, Am Hubland, D-97074 W\"{u}rzburg, Germany}
\affiliation{W\"{u}rzburg-Dresden Cluster of Excellence ct.qmat, Germany}

\author{Bastian~Rubrecht}
\affiliation{Leibniz Institut f\"{u}r Festk\"{o}rper- und Werkstoffforschung (IFW) Dresden, Helmholtzstra\ss{}e 20, D-01069 Dresden, Germany}

\author{Jorge~I.~Facio}
\affiliation{Leibniz Institut f\"{u}r Festk\"{o}rper- und Werkstoffforschung (IFW) Dresden, Helmholtzstra\ss{}e 20, D-01069 Dresden, Germany}
\affiliation{Centro At\'omico Bariloche, Instituto de Nanociencia y Nanotecnolog\'ia (CNEA-CONICET) and Instituto Balseiro. Av. Bustillo 9500, Bariloche (8400), Argentina.}

\author{Volodymyr~B.~Zabolotnyy}
\affiliation{Physikalisches Institut (EP-IV), Fakult\"{a}t f\"{u}r Physik und Astronomie, Universit\"{a}t W\"{u}rzburg, Am Hubland, D-97074 W\"{u}rzburg, Germany}
\affiliation{W\"{u}rzburg-Dresden Cluster of Excellence ct.qmat, Germany}

\author{Laura~T.~Corredor}
\affiliation{Leibniz Institut f\"{u}r Festk\"{o}rper- und Werkstoffforschung (IFW) Dresden, Helmholtzstra\ss{}e 20, D-01069 Dresden, Germany}

\author{Laura~C.~Folkers}
\affiliation{Institut f\"{u}r Festk\"{o}rper- und Materialphysik, Technische Universit\"{a}t Dresden, 01062 Dresden, Germany}
\affiliation{W\"{u}rzburg-Dresden Cluster of Excellence ct.qmat, Germany}

\author{Ekaterina~Kochetkova}
\affiliation{Leibniz Institut f\"{u}r Festk\"{o}rper- und Werkstoffforschung (IFW) Dresden, Helmholtzstra\ss{}e 20, D-01069 Dresden, Germany}

\author{Thiago~R.~F.~Peixoto}
\affiliation{Physikalisches Institut (EP-VII), Fakult\"{a}t f\"{u}r Physik und Astronomie, Universit\"{a}t W\"{u}rzburg, Am Hubland, D-97074 W\"{u}rzburg, Germany}
\affiliation{W\"{u}rzburg-Dresden Cluster of Excellence ct.qmat, Germany}

\author{Philipp~Kagerer}
\affiliation{Physikalisches Institut (EP-VII), Fakult\"{a}t f\"{u}r Physik und Astronomie, Universit\"{a}t W\"{u}rzburg, Am Hubland, D-97074 W\"{u}rzburg, Germany}	
\affiliation{W\"{u}rzburg-Dresden Cluster of Excellence ct.qmat, Germany}	

\author{Simon~Heinze}
\affiliation{Institute for Theoretical Physics, Heidelberg University, Philosophenweg 19, 69120 Heidelberg, Germany}
	
\author{Hendrik~Bentmann}
\affiliation{Physikalisches Institut (EP-VII), Fakult\"{a}t f\"{u}r Physik und Astronomie, Universit\"{a}t W\"{u}rzburg, Am Hubland, D-97074 W\"{u}rzburg, Germany}	
\affiliation{W\"{u}rzburg-Dresden Cluster of Excellence ct.qmat, Germany}	
	
\author{Robert~J.~Green}
\affiliation{Department of Physics and Astronomy and Stewart Blusson Quantum Matter Institute, University of British Columbia, Vancouver, BC V6T 1Z4, Canada}
\affiliation{ Department of Physics and Engineering Physics, University of Saskatchewan, SK S7N 5E2 Saskatoon, Canada}

\author{Pierluigi~Gargiani}
\affiliation{ALBA Synchrotron Light Source, E-08290 Cerdanyola del Vall\`{e}s, Barcelona, Spain}

\author{Manuel~Valvidares}
\affiliation{ALBA Synchrotron Light Source, E-08290 Cerdanyola del Vall\`{e}s, Barcelona, Spain}	

\author{Eugen~Weschke}
\affiliation{Helmholtz-Zentrum Berlin f\"{u}r Materialien und Energie, Albert-Einstein-Stra\ss{}e 15, D-12489 Berlin, Germany}

\author{Maurits~W.~Haverkort}
\affiliation{Institute for Theoretical Physics, Heidelberg University, Philosophenweg 19, 69120 Heidelberg, Germany}

\author{Friedrich~Reinert}
\affiliation{Physikalisches Institut (EP-VII), Fakult\"{a}t f\"{u}r Physik und Astronomie, Universit\"{a}t W\"{u}rzburg, Am Hubland, D-97074 W\"{u}rzburg, Germany}
\affiliation{W\"{u}rzburg-Dresden Cluster of Excellence ct.qmat, Germany}	

\author{Jeroen~van~den~Brink}
\affiliation{Leibniz Institut f\"{u}r Festk\"{o}rper- und Werkstoffforschung (IFW) Dresden, Helmholtzstra\ss{}e 20, D-01069 Dresden, Germany}
\affiliation{Institut f\"{u}r Theoretische Physik, Technische Universit\"{a}t Dresden, D-01062 Dresden, Germany}
\affiliation{W\"{u}rzburg-Dresden Cluster of Excellence ct.qmat, Germany}

\author{Bernd~B\"{u}chner}
\affiliation{Leibniz Institut f\"{u}r Festk\"{o}rper- und Werkstoffforschung (IFW) Dresden, Helmholtzstra\ss{}e 20, D-01069 Dresden, Germany}
\affiliation{Institut f\"{u}r Festk\"{o}rper- und Materialphysik, Technische Universit\"{a}t Dresden, 01062 Dresden, Germany}
\affiliation{W\"{u}rzburg-Dresden Cluster of Excellence ct.qmat, Germany}

\author{Anja~U.~B.~Wolter}
\affiliation{Leibniz Institut f\"{u}r Festk\"{o}rper- und Werkstoffforschung (IFW) Dresden, Helmholtzstra\ss{}e 20, D-01069 Dresden, Germany}
\affiliation{W\"{u}rzburg-Dresden Cluster of Excellence ct.qmat, Germany}

\author{Anna~Isaeva}
\email[Corresponding address: ]{a.isaeva@uva.nl}
\affiliation{Van der Waals-Zeeman Institute, Department of Physics and Astronomy, University of Amsterdam, Science Park 904, 1098 XH Amsterdam, The Netherlands}
\affiliation{Leibniz Institut f\"{u}r Festk\"{o}rper- und Werkstoffforschung (IFW) Dresden, Helmholtzstra\ss{}e 20, D-01069 Dresden, Germany}

\author{Vladimir~Hinkov}
\email[Corresponding address: ]{hinkov@physik.uni-wuerzburg.de}
\affiliation{Physikalisches Institut (EP-IV), Fakult\"{a}t f\"{u}r Physik und Astronomie, Universit\"{a}t W\"{u}rzburg, Am Hubland, D-97074 W\"{u}rzburg, Germany}
\affiliation{W\"{u}rzburg-Dresden Cluster of Excellence ct.qmat, Germany}

\date{\today}

\renewcommand{\figurename}{FIG.}

\begin{abstract}
The recent realizations of the quantum anomalous Hall effect (QAHE) in MnBi$_2$Te$_4$  and MnBi$_4$Te$_7$ benchmark the (MnBi$_2$Te$_4$)(Bi$_2$Te$_3$)$_n$ family as a promising hotbed for further QAHE improvements. The family owes its potential to its ferromagnetically (FM) ordered MnBi$_2$Te$_4$ septuple layers (SL). However, the QAHE realization is complicated in MnBi$_2$Te$_4$ and MnBi$_4$Te$_7$ due to the substantial antiferromagnetic (AFM) coupling between the SL. An FM state, advantageous for the QAHE, can be stabilized by interlacing the SL with an increasing number $n$ of Bi$_2$Te$_3$ layers. However, the mechanisms driving the FM state and the number of necessary QLs are not understood, and the surface magnetism remains obscure. Here, we demonstrate robust FM properties in MnBi$_6$Te$_{10}$ ($n=2$) with $T_c\approx12\,\text{K}$ and establish their origin in the Mn/Bi intermixing phenomenon by a combined experimental and theoretical study. Our measurements reveal a magnetically intact surface with a large magnetic moment, and with FM properties similar to the bulk. Our investigation thus consolidates the MnBi$_6$Te$_{10}$ system as perspective for the QAHE at elevated temperatures.
\end{abstract}

\maketitle

\section{Introduction}

Theory provides a seemingly straightforward avenue towards novel quantum effects such as the quantum anomalous Hall (QAH) effect \cite{KOU201534, Annurev_He, QAHCr, QAHV, PhysRevLett.101.146802} and axion electrodynamics \cite{Li2010, PhysRevLett.118.246801,PhysRevLett.120.056801,PhysRevLett.122.206401}, namely to induce a long-range ferromagnetic (FM) order in topological insulators (TI) \cite{HasanTI, QiTISC}. The vision of observing Majorana fermions and implementing topological qubits at superconductor/QAH insulator interfaces \cite{He17_Science357_294}, ultra low-power electronics \cite{lowPowerElectronics} and applications in spintronics \cite{spintronics} has ignited substantial experimental efforts in this direction. Yet, hitherto the QAH effect (QAHE) has only been demonstrated in the (sub-) kelvin range~\cite{QAHCr, QAHV,PhysRevLett.114.187201}. The experimental realization of the QAHE is complicated by several simultaneous requirements to a candidate system: The Dirac point (DP) of the parent TI should be well within its bulk band gap; the chemical potential has to be tuned to the DP; the introduced magnetic subsystem should lead to a substantial surface ferromagnetism to open a large exchange gap at the DP;  and the material's bulk should remain insulating. 

\begin{figure*}
    \centering
    \includegraphics[width=0.83\textwidth]{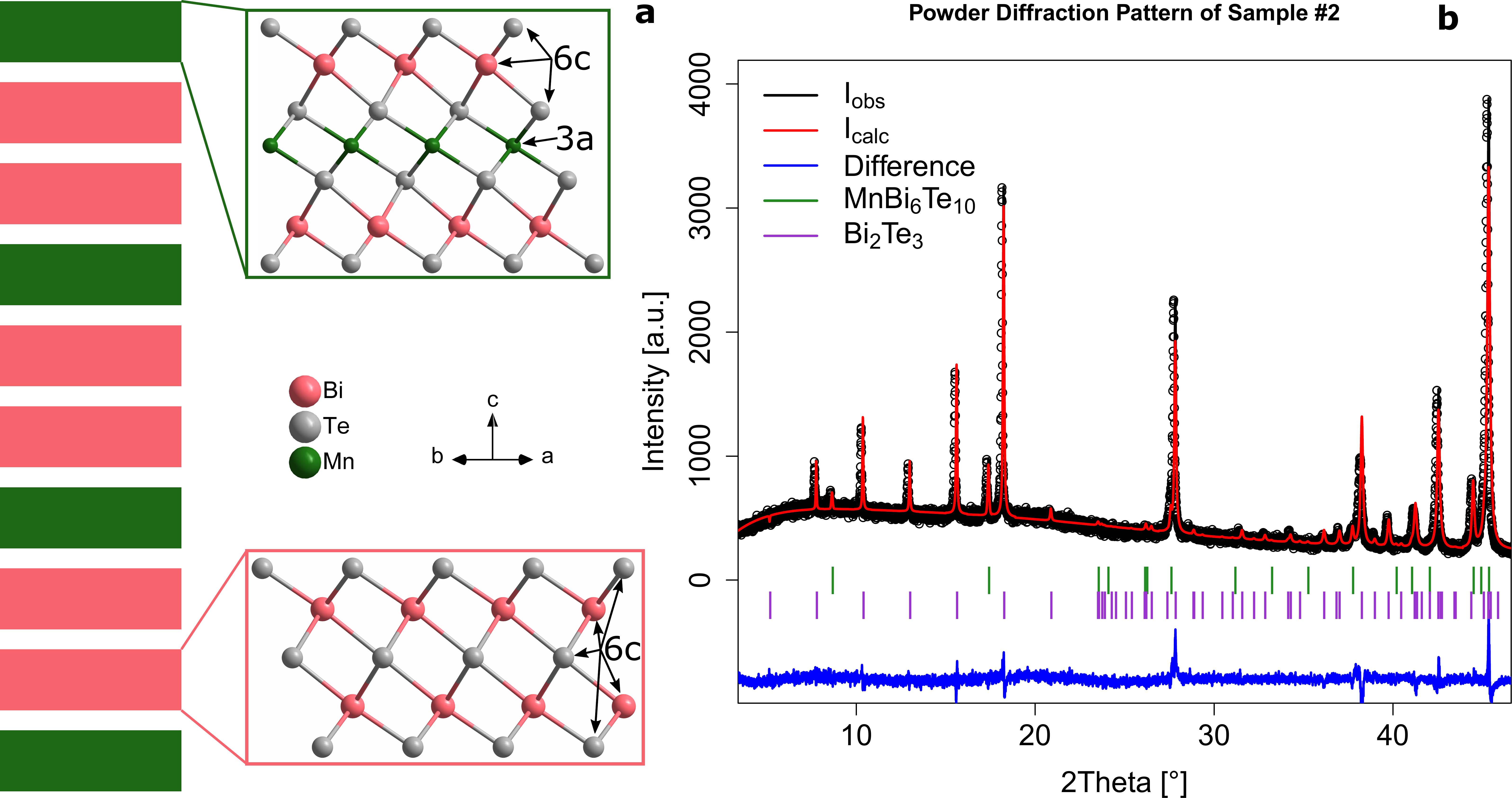}
    \caption{Crystal structure properties. a) The unit cell of MnBi$_6$Te$_{10}$ is sketched by slabs of red and green boxes, where green indicates a septuple layer and red indicates a quintuple layer. In the expanded views we show the atomic structure. The QL and SL are interleaved by van der Waals gaps. b) Experimental (black) and refined by Rietveld method (red) powder X-ray diffraction pattern of the sample $\#$2 in the 2$\theta$ range $5-45 \degree$. For the full 2$\theta$ range, see~Fig.~S1. The difference curve is shown in blue ($R_p = 0.055$, $wR_p = 0.071$, GoF = 1.48). A small fraction of Bi$_2$Te$_3$ comprises 7~wt. \% ($R\bar{3}m$, $a = 4.3797(4)$ {\AA}, $c = 30.4965(7)$ {\AA}, $R_{obs} = 0.094, wR_{obs} = 0.096, R_{all} = 0.109$). The main phase is refined with the overall Mn$_{0.8}$Bi$_{6.2}$Te$_{10}$ composition ($R\bar{3}m$, $a = 4.3667(2)$ \AA, $c = 101.869(4)$ \AA, $R_{obs} = 0.079, wR_{obs} = 0.074, R_{all} = 0.104$).}
    \label{fig:stack}
\end{figure*}

The first materials to exhibit the QAHE were extrinsically doped (V/Cr)$_x$(Bi,Sb)$_{2-x}$Te$_3$, which consist of van-der-Waals coupled quintuple layers (QL, see Fig.~\ref{fig:stack}). However, band engineering by tuning the Bi/Sb ratio does not move the DP sufficiently above the valence band \cite{Li16_SciRep6_32732}, V/Cr impurity bands overlap with the alleged exchange gap \cite{Peixoto20_npjQM5_87}  and residual bulk conductance destroys quantization with increasing temperature \cite{Fijalkowsi21_NatComm12_5599}. As a result, the QAHE is stable only below $T_\text{QAH}=20\,\text{mK}$.

The intrinsic magnetic topological insulators (MnBi$_2$Te$_4$)(Bi$_2$Te$_3$)$_n$ (MBT$_n$, $n=0-4$), whose functional constituents are (MnBi$_2$Te$_4$) septuple layers (SLs) with the central sheet of FM-ordered Mn atoms, separated by $n$ Bi$_2$Te$_3$ QLs, offer several advantages.  Whereas for QL termination, angle-resolved photoemission spectroscopy (ARPES) measurements on MBT$_1$ and MBT$_2$ yield similar results to  (Bi,Sb)$_{2}$Te$_3$, without a discernible DP, there is a DP within the bulk gap for the SL termination~\cite{vidal2020orbital}. Also, the topmost sheet of the ferromagnetically arranged Mn moments should strongly couple with the topological surface states (TSS), albeit a thorough spectroscopic investigation of the surface magnetism is still pending. As a result, a much higher $T_\text{QAH}=1.4\,\text{K}$ is achieved in MnBi$_2$Te$_4$ \cite{Deng895}. This is despite the fact that MnBi$_2$Te$_4$ is suboptimal due to its antiferromagnetic (AFM) order ($T_\text{N}=24\,\text{K}$) and a complex layer-number dependence of the quantization effects, with an odd number of SLs required to realize the QAHE~\cite{Deng895}. 

Yet, the potential of the other MBT$_n$ for a further substantial increase of $T_\text{QAH}$ is strong: Increasing $n$ weakens the interlayer AFM coupling so that FM properties gradually develop. Indeed, most studies report a complex metamagnetic behavior in MnBi$_4$Te$_7$ and MnBi$_6$Te$_{10}$~ \cite{PhysRevLett.124.197201, Hueaba4275, Klimovskikh2020, Xie2020, PhysRevB.102.035144,PhysRevX.9.041065, PhysRevB.102.045130},  but a clear FM state only for $n\geq3$ \cite{Klimovskikh2020,Hueaba4275,PhysRevX.11.011039}. Already metamagnetic MnBi$_4$Te$_7$ hosts the QAHE up to several degree kelvin in the bulk regime~\cite{Deng2021}. This experimental realization of a QAHE device out of a bulk crystal required technically challenging but feasible efforts to attain the charge neutrality condition~\cite{Deng2021}. Consequently, envisioning this clear, technical realization path once an appropriate bulk crystal exists, we focus here on the next obvious step, namely to strengthen the FM properties of MBT$_n$. Our synthesis endeavors culminate in the robust FM order in MnBi$_6$Te$_{10}$ (i.e. already for $n=2$), an intrinsic magnetic TI~\cite{vidal2020orbital} and QAHE candidate~\cite{Otrokov_2017}.

We confirm the FM state both in the bulk and on the surface of MnBi$_6$Te$_{10}$ crystals by using bulk-sensitive superconducting quantum interference device (SQUID) magnetometry and surface-sensitive x-ray magnetic circular dichroism (XMCD). The clear FM characteristics seemingly contradict the weak AFM coupling anticipated by our density functional theory (DFT) calculations for the atomically ordered compound. This disagreement is resolved by including the experimentally determined Mn substoichiometry and Mn/Bi site intermixing into account. Our calculations pinpoint that the magnetic coupling can be tuned towards ferromagnetism by appropriate intermixing already in MnBi$_4$Te$_7$ and even MnBi$_2$Te$_4$. Considering the intermixing patterns in our MnBi$_6$Te$_{10}$ samples and those reported showing no ferromagnetism, we rationalize their differing magnetic behavior. Our results demonstrate  that carefully engineered intermixing can accomplish a robust FM order and, therefore, is the key towards enhanced QAHE properties in the MBT$_n$ family of intrinsic magnetic topological insulators.

\section{Results}
 
 \subsection{Crystal growth and structure refinement}
 
  \label{sec:growth}
 
 MnBi$_{6}$Te$_{10}$ crystals were grown by slow crystallization from a melt (see Sec. \ref{sec:METHODS}). Besides MnBi$_6$Te$_{10}$, the obtained ingot contained  admixtures of Bi$_2$Te$_3$ and MnTe$_2$ (see Fig.~S1 in the Supporting Information). Observing side phases fully agrees with our earlier studies of MnBi$_6$Te$_{10}$ melting and decomposition by differential scanning calorimetry~\cite{C9TC00979E}. Their occurrence can be related to crystal growth being a competitive process between MnBi$_6$Te$_{10}$, MnBi$_8$Te$_{13}$ and Bi$_2$Te$_{3}$, all having nearly the same crystallization temperatures.
 
 A series of EDX (energy-dispersive x-ray spectroscopy) point measurements on individual crystals extracted from the ingot demonstrated a compositional range between Mn: $5.0$, Bi: $36.6$, Te: $58.4$ and  Mn: $4.2$, Bi: $37.1$, Te: $58.7$ (in at.~$\%$). Our samples were thus consistently more Mn-deficient than expected from the nominal chemical formula MnBi$_6$Te$_{10}$ of the atomically ordered material (Mn: $5.9$, Bi: $35.3$; Te: $58.8$). Again, this echoes our earlier published single-crystal structure refinement of  Mn$_{0.73(4)}$Bi$_{6.18(2)}$Te$_{10}$ by x-ray diffraction~\cite{C9TC00979E}, where we systematically showed that Mn-substoichiometry is determined by the Mn/Bi intermixing. Both features are also present in Mn$_{0.85}$Bi$_{2.10}$Te$_{4}$~\cite{Zeugner2019} and Mn$_{0.75}$Bi$_{4.17}$Te$_{7}$~\cite{PhysRevX.9.041065}. To facilitate perception, we denote our samples as MnBi$_6$Te$_{10}$ in the following text, keeping in mind that they are in fact substoichiometric.
 
  The present study was performed on four individual Mn-deficient MnBi$_6$Te$_{10}$ crystals (denoted as Sample $\#1-\#4$ henceforward; for their chemical compositions (EDX) see Fig.~S2).
  Powder x-ray diffraction (PXRD) measurements, which required grinding the crystals to a homogeneous powder, were conducted after all other measurements had been finalized, in order to elucidate the underlying intermixing phenomenon. We confirmed that all four samples  exhibit the crystal lattice of MnBi$_6$Te$_{10}$ with a sequence of one SL and two QLs  (Fig.~\ref{fig:stack}a) plus notable cation antisite disorder.  MnBi$_6$Te$_{10}$ constituted the main phase as per Rietveld method and we established a firm link between the Mn content as found by EDX and the underlying crystal lattice of MnBi$_6$Te$_{10}$ in our samples. 
  
  This approach is exemplified on Sample~$\#2$ (see~Fig.~\ref{fig:stack}b and more procedural details in the Supporting Information sec. I). We confirmed that sample~$\#2$ was Mn$_{1-x}$Bi$_{6+x}$Te$_{10}$ ($x \approx 0.20-0.25$) which crystallized in the rhombohedral space group $R\bar{3}m$ (No.~166) with the unit cell lattice parameters $a=4.36778(8)$ {\AA} and $c=101.8326(6)$ {\AA}. To stabilize a further Rietveld refinement, the EDX compositions (\textit{e.g.} Mn$_{0.76}$Bi$_{6.24}$Te$_{10}$ or Mn$_{0.8}$Bi$_{6.2}$Te$_{10}$) were introduced as constraints (see Supporting Information sec. I). When cation Mn/Bi intermixing was allowed in the refinement, the reliability factors $R_{all}$ and $R_{obs}$ dropped down significantly, confirming that this phenomenon was undoubtedly present in the structure. Due to very low sample mass (1--2~mg), the acquired powder diffraction data did not allow us to settle in for just one particular intermixing model with a statistically unequivocal quantification. The refined Mn content is also strongly dependent on whether cation vacancies are allowed in the refinement. We opted for a structural solution without voids in the $3a$ and $6c$ positions. Despite the outlined uncertainties, all tested models with various composition constraints have in common that: 1) the Mn:Bi ratio in the 3{\it a} position in the center of an SL is close to 56:44; 2) the outer cation site of an SL (6{\it c}) contains up to 2~\% Mn; 3) the QL always accommodates some Mn (2--7~\% Mn) in the  $6c$ cation sites. The presence of Mn in all cation positions accords with our earlier reported refinement on Mn$_{0.81}$Bi$_{6.13}$Te$_{10}$ single crystals~\cite{C9TC00979E} and is in contrast to the findings of Klimovskikh \textit{et al.}~\cite{Klimovskikh2020}. Such subtle variations in intermixing patterns can dramatically impact the magnetic properties, as witnessed in the next subsection.

 \subsection{Bulk magnetometry}
 
  \label{sec:SQUID}
 
  \begin{figure*}
	\centering
	\includegraphics[width=0.85\textwidth]{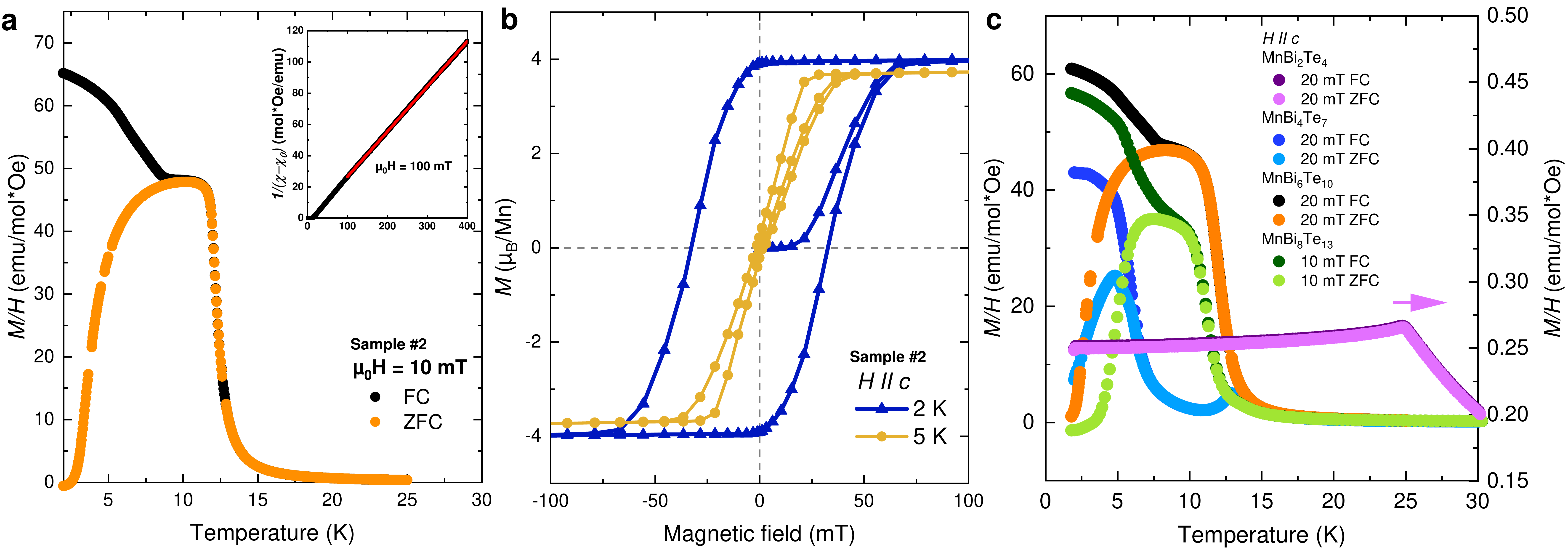}
	\caption{SQUID magnetometry measurements.  (a) Temperature-dependent normalized magnetization $M/H$ of sample \#2 with ZFC (orange symbols) and (FC (black symbols) protocols in an out-of-plane applied magnetic field of $10\,\text{mT}$. The inset shows the inverse magnetic susceptibility in a magnetic field of $100\,\text{mT}$ together with a modified Curie-Weiss fit $\chi(T)=\chi_0+C/(T-\Theta_\text{CW})$ of the data above $100\,\text{K}$ (red solid line); for details see the Methods section. (b) Field-dependent magnetization of sample \#2 measured in an out-of-plane applied magnetic field at $T=2\,\text{K}$ and $5\,\text{K}$. No demagnetization correction was applied, and the magnetization was normalized to the Mn content obtained by EDX. (c) Temperature dependence of the normalized magnetization of analogously synthesized samples of the MBT$_n$ family for ($n=0,1,2,3$).}
	\label{fig:SQUID}
\end{figure*}
 
 Fig.~\ref{fig:SQUID}a shows the field-cooled (FC) and zero-field cooled (ZFC) normalized magnetization of sample $\#2$ in an out-of-plane magnetic field of $10\,\text{mT}$. A phase transition into a long-range magnetically ordered state is observed at $T_c = 12.0\,\text{K}$, determined by the inflection point, together with a notable FC/ZFC splitting around $10\,\text{K}$. These observations point towards a ferromagnetic alignment of the Mn spins in our MnBi$_6$Te$_{10}$ samples and contrast with the antiferromagnetic transition at $T_\text{N}\sim11\,\text{K}$ so far reported for the nominal MnBi$_{6}$Te$_{10}$ composition \cite{Klimovskikh2020,PhysRevB.102.035144,PhysRevB.100.155144,PhysRevMaterials.4.054202, PhysRevB.102.045130}. Our Curie-Weiss analysis in the temperature regime 100--400\,K (see inset of Fig.~\ref{fig:SQUID}a and Sec.~\ref{sec:METHODS}) yields an effective moment of $m_\text{eff} = 5.8 \pm 0.1~ \mu_\text{B}/\text{Mn}$ in close agreement with the  value $m_{\textrm{eff}} = 5.67 \mu_\textrm{B}$ calculated by multiplet ligand-field theory (MLFT) (sec.~\ref{sec:MLFT}). The uniformity of all four MnBi$_6$Te$_{10}$ crystals is strongly supported by the nearly identical SQUID magnetometry curves (see Fig. S4),  with transition temperatures that vary by only $0.1\,\text{K}$.

The magnetization curves  $M(H)$ in Fig.~\ref{fig:SQUID}b show clear FM loop openings, with a coercive field of $\mu_0H_c \sim32\,\text{mT}$ at $T=2\,\text{K}$, and a finite remanent moment of $m_\text{SQ}^\text{tot}=(3.9\pm0.2)\mu_\text{B}/\text{Mn}$ at zero magnetic field. The moment at $0.15\,\text{T}$ is $m_\text{SQ}^\text{tot}=(4.2\pm0.2)\mu_\text{B}/\text{Mn}$.

It is furthermore interesting to compare our results to analogously synthesized samples of the MBT$_n$ family (Fig.~\ref{fig:SQUID}c). We observe a noteworthy trend as the number of quintuple Bi$_2$Te$_{3}$ layers $n$ increases: MnBi$_2$Te$_{4}$ ($n=0$) has a clear A-type AFM structure, whereas MnBi$_4$Te$_{7}$ ($n=1$) exhibits a more complex behavior, in which robust low-temperature metamagnetic properties are established, which were shown to result from the competition between the uniaxial anisotropy $K$ and the still sizable interlayer AFM interaction $J$ \cite{PhysRevLett.124.197201}. Finally, in MnBi$_6$Te$_{10}$ ($n=2$), as well as in MnBi$_8$Te$_{13}$ ($n=3$), the FM properties clearly dominate, with FM order at the significant temperatures of $T_c = 12\,\text{K}$ and $10\,\text{K}$, respectively. Consistent with this observation, the spin-flop transition found for MnBi$_2$Te$_4$ and MnBi$_4$Te$_7$ at fields of $3.5~\text{T}$ \cite{Otrokov2019,Wueaax9989,Zeugner2019} and $0.1$--$0.3~\text{T}$ \cite{PhysRevX.9.041065, Klimovskikh2020,Hu2020,PhysRevB.100.155144}, respectively, is absent in MnBi$_6$Te$_{10}$, and a magnetic moment of more than 4$\mu_\textrm{B}$ is observed already above 80~mT in the latter after a ZFC procedure.

\subsection{Bulk DFT (GGA+$U$) calculations}
\label{sec:DFT-GGA}

We have first performed fully relativistic DFT calculations based on the Generalized Gradient Approximation (GGA)$+U$ \cite{perdew1996generalized} for MnBi$_6$Te$_{10}$ neglecting the intermixing. For the interaction parameters, we have used the Slater integrals in Table~\ref{table:Slater} for the initial state. The results of total energy calculations for the A-type AFM configuration favor the out-of-plane over the in-plane magnetization by $\sim 0.4$\,meV per Mn. Additional calculations indicate that the A-type AFM configuration has a lower energy than the FM configuration. However, the small magnitude of the difference, $\sim 0.04\,$meV per Mn, naturally suggests that other mechanisms such as Mn/Bi intermixing may well be relevant for the magnetic ground state.

\begin{figure*}[t]
	\centering
\includegraphics[width=\textwidth]{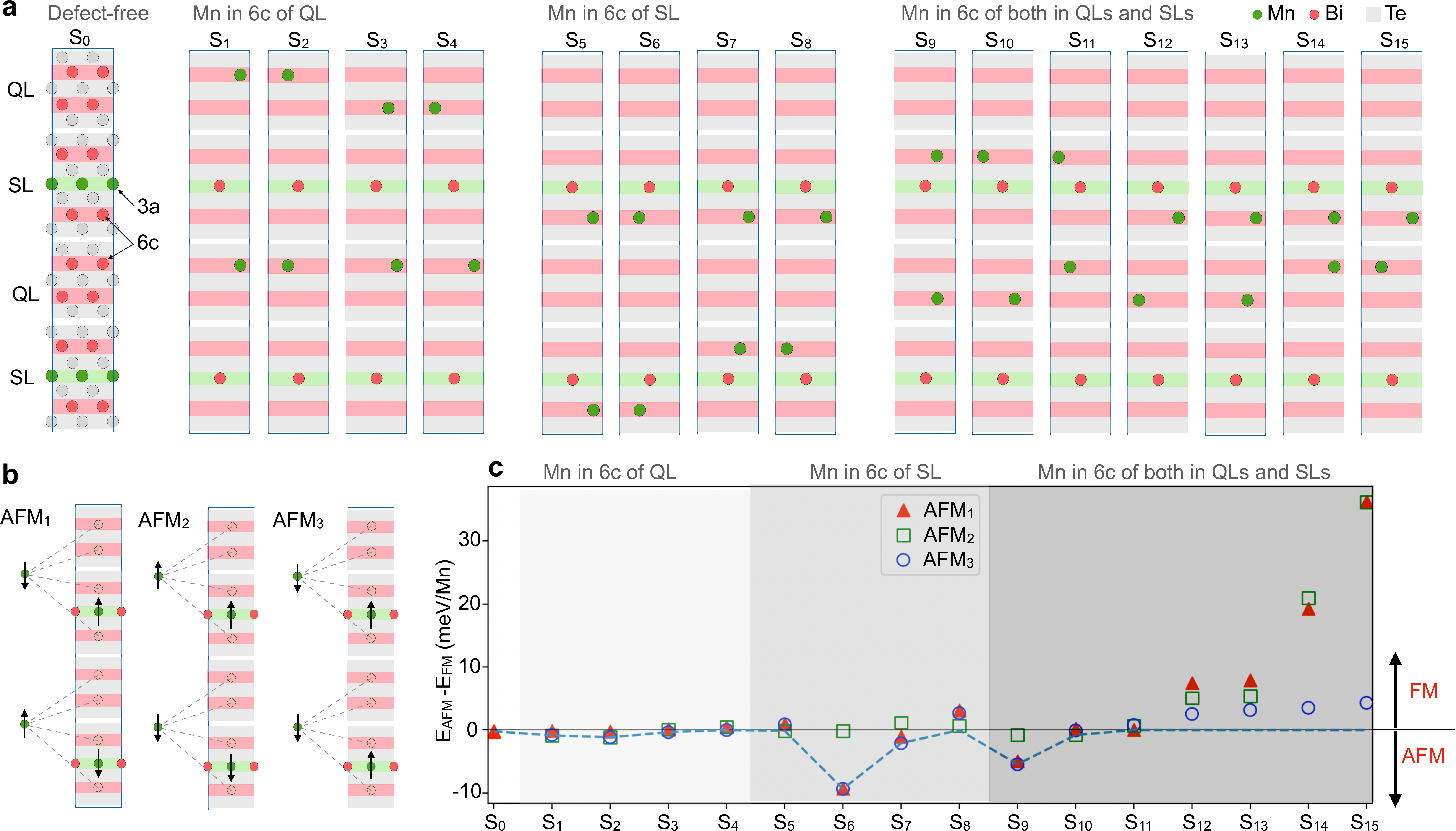} 
	\caption{Crystal- and magnetic-structure models and their DFT energies. (a) Schematics of the structure models with various Mn/Bi intermixing scenarios. S$_0$ is the defect-free case while the panels from S$_1$ to S$_{15}$ visualize the structural differences to S$_0$. Models S$_1$ to S$_4$: antisite Mn (green) is in the $6c$ site of the QLs; models S$_5$ to S$_8$: antisite Mn is in the $6c$ site of the SLs; models S$_{9}$ to S$_{15}$: antisite Mn is in the $6c$ positions of both QLs and SLs.  (b) Schematics of the constructed magnetic arrangements. Each model has a distinct order between the Mn magnetic moments in the $3a$ and $6c$ positions which are conditioned by the respective structure model in (a). Ferromagnetic order with an out-of-plane orientation of the moments is assumed within each atomic layer. (c) Total energy difference between an antiferromagnetic ordering model (1, 2 or 3) of each structure model (S$_0$ to S$_{15}$) and the respective fully ferromagnetic configuration, as obtained from the scalar-relativistic DFT calculations. The dashed line follows the ground state energy, zero corresponding to the FM phase.}
	\label{fig:theory}
\end{figure*}

Taking into account the Mn/Bi intermixing for MnBi$_6$Te$_{10}$, with its lattice parameter $c>100\,\text{\AA}$, would require a prohibitively long computational time. Instead, here we aim to learn the effects of Mn/Bi intermixing on magnetism via the simpler models of MnBi$_2$Te$_4$ and MnBi$_4$Te$_7$. The latter case is more representative of MnBi$_6$Te$_{10}$, since it contains both SLs and QLs, and is discussed in detail below, while the former is presented in the Supporting Information (Sec.~IIIA). Here, we emphasize one conclusion about MnBi$_2$Te$_4$: even though its defect-free form has the strongest AFM coupling between Mn in the consecutive SLs, based on our calculations, intermixing  can induce the FM order between the SLs even in this compound (Fig.~S5 in the Supporting Information). Hence, the emergence of strong out-of-plane FM correlations due to the intermixing is likely to be universally present in the MBT$_n$ family, including  MnBi$_6$Te$_{10}$.

As $n$ increases, the possibilities for intermixing patterns naturally become larger as antisite Mn atoms can be located in the $6c$ positions (occupied by Bi in the defect-free case) of both the SLs and the QLs. 
MnBi$_4$Te$_7$ provides the minimal framework to explore whether this enlarged configuration space can yield variations in the experimentally observed ground states. We have performed scalar-relativistic calculations for various intermixing patterns and magnetic orders in a $2\times1\times2$ supercell of MnBi$_4$Te$_7$ (Fig.~\ref{fig:theory}). In addition to the defect-free case (S$_0$), we construct models (S$_1$ to S$_{15}$) with different Mn/Bi antisite defects, all of them globally stoichiometric and having a 50\% fraction of Bi atoms in the $3a$ Wyckoff site. Notice that this concentration is close to the outcome of our Rietveld refinements ($\sim44\%$). The models differ in the  positions occupied by the antisite Mn atoms and can be classified into three categories. In the first category, the antisite Mn atoms reside in the $6c$ position of {\it only} the QLs (S$_1$ to S$_4$). Similarly, in the second category antisite Mn occupy the $6c$ positions of {\it only} the SLs (S$_5$ to S$_8$). In the third category, the Mn atoms are distributed over the $6c$ positions of {\it both} the QLs and the SLs (S$_9$ to S$_{15}$).

For each structure model, we consider four possible magnetic arrangements: a fully spin-polarized FM order and three different AFM models sketched in Fig.~\ref{fig:theory}b. They all have in common that the Mn moments order FM within any given atomic layer, but vary in the magnetic couplings between the adjacent atomic layers along the stack. In the  AFM$_1$ and AFM$_2$ models, the Mn spins in the two consequent  3$a$ positions are  oppositely coupled. The coupling between the $3a$ site and all intermixed Mn neighbours in the $6c$ site(s) is either AFM (AFM$_1$) or FM (AFM$_2$), respectively. The AFM$_3$ model realises parallel spin arrangement in the $3a$ sites, while they couple AFM with the Mn defects in all $6c$ positions.

Fig.~\ref{fig:theory}c discerns what is an energetically favorable magnetic arrangement for each considered structure model of MnBi$_4$Te$_7$ as compared to the fully spin-polarized FM state. For a given model, if any AFM model obeys $E_{AFM}-E_{FM}<0$, we conclude that antiferromagnetism is preferred. On the other hand, if all AFM models fulfill $E_{AFM}-E_{FM}>0$, we define the fully FM phase as the ground state.

A clear trend in the magnetic order as a function of the underlying Mn/Bi intermixing pattern can be established. All but one models of the first two categories, where the intermixed Mn cations occupy {\it either} the SLs {\it or} the QLs, show an AFM configuration as the lower energy state. 
This preference reverts markedly if the antisite Mn distributes over the $6c$ positions of {\it both} the QLs and the SLs: five out of the seven constructed structure configurations of this category prefer the FM phase. A closer look reveals that the preference for the FM state is particularly prominent in those structure models (S$_{12}$ to S$_{15}$), in which the Mn cations occupy the nearby $6c$ positions not separated by the $3a$ positions -- namely, when continuous magnetic exchange pathways exist between the antisite Mn ions.

These results establish a strong correlation between the magnetic structure and the Mn distribution along the stacking direction. When an antisite Mn is located only in one of the two $6c$ positions of the QLs -- a situation experimentally found in Ref.~\cite{Klimovskikh2020} -- our calculations suggest the prevalence of an AFM order. When Mn distributes both in the $6c$ of the QLs and SLs, which is the case of our samples according to our structure refinements, our calculations identify the FM phase as the ground state.

\subsection{XAS and XMCD Data}

\label{sec:XASXMCD}

To study the surface magnetic properties, we have performed x-ray absorption spectroscopy (XAS) measurements in the total electron yield (TEY) mode, which is element specific and has a probing depth on the nanometer scale. Measurements at the Bi $N_{4,5}$ edges (Fig. S7b) exhibit no XMCD. More interestingly, there is no XMCD at the Te $M_{4,5}$ edges either (Fig. S7a), which is in contrast to results in the closely related V- and Cr-doped (Bi,Sb)$_2$Te$_3$ \cite{Tcakaev,PhysRevB.97.155429, PhysRevB.99.144413, Ye2015}, and which might indicate differences in the magnetic interactions of both compounds. 

\begin{figure}[b]
	\centering
	\includegraphics[width=1.0\columnwidth]{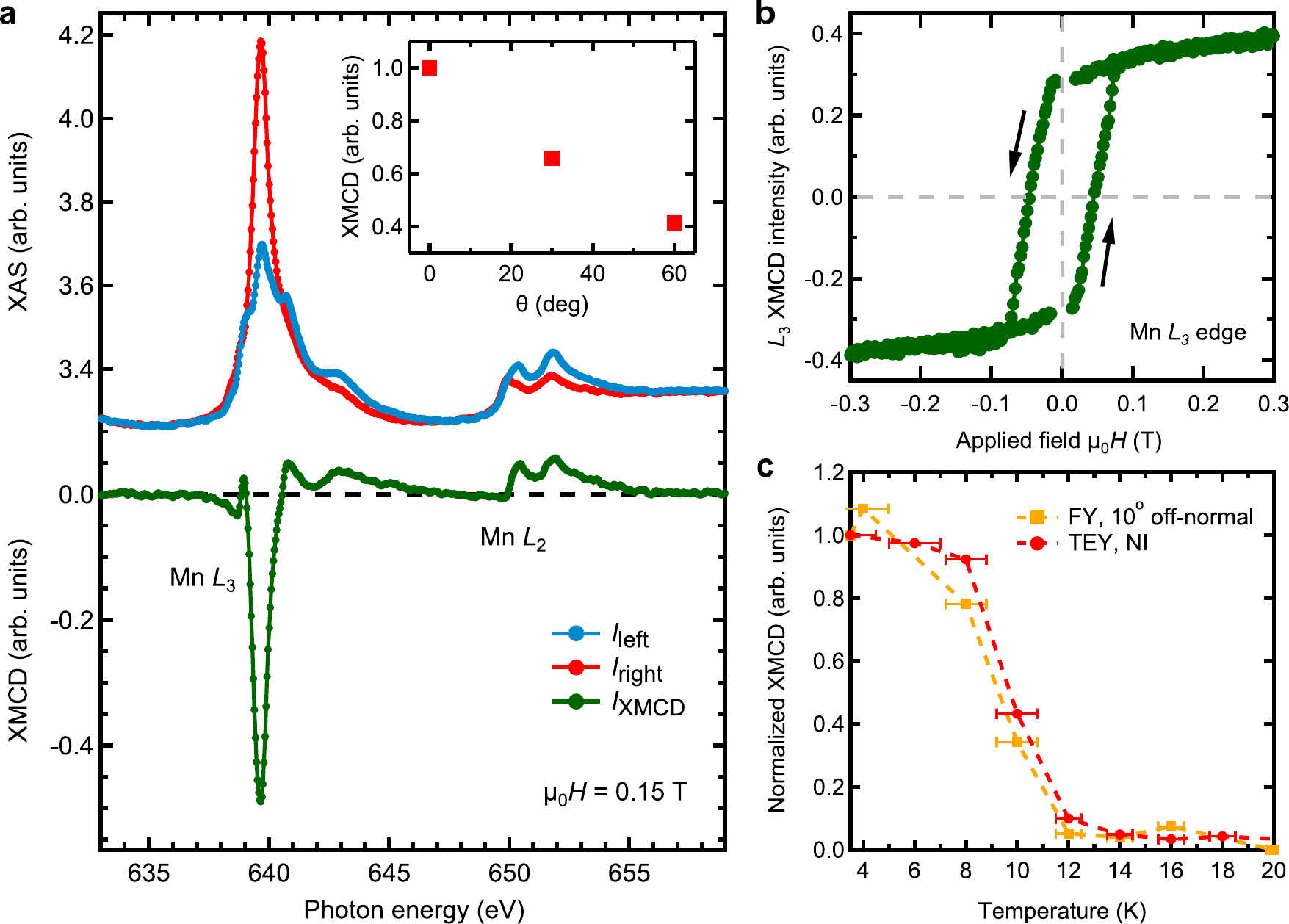}
	\caption{X-ray spectroscopy data. (a)  Mn $L_{2,3}$ edge XAS data for sample $\#4$ obtained with left ($I_\text{left}$, blue) and right ($I_\text{right}$, red) circularly polarized light in normal incidence at $T\approx3.5~\text{K}$ in a magnetic field of $0.15\,\text{T}$. The corresponding XMCD signal $I_\text{XMCD}=I_\text{left}-I_\text{right}$ is plotted below in green. The inset shows the angular dependence of the normalized remanent XMCD signal. (b) Magnetization curve of sample $\#4$ ($I_\text{right}-I_\text{left}$) at $T \approx 3.5\,\text{K}$, obtained as the Mn $L_3$ edge XMCD signal normalized by the XAS signal. (c) Temperature dependence of the remanent XMCD signal for sample $\#2$ at the Mn $L_3$ edge measured at normal incidence in TEY mode (red) and $10^\circ$ off normal incidence in FY mode (orange).}
	\label{fig:XMCD}
\end{figure}

Next we focus on the Mn $L_{2,3}$ edge. Due to the shallow escape depth, the topmost SL contributes the most to the signal. However, even for SL termination, the FM Mn sheet is buried about $0.55\,\text{nm}$ below the surface, and significantly more for QL termination. Therefore, probing depth effects have to be considered, when interpreting the ordered magnetic moments obtained with XMCD (see secs.~\ref{sec:SumRules} and \ref{sec:discussion}). Fig.~\ref{fig:XMCD}a compares XAS spectra measured with x-rays of opposite circular polarization at $T\approx3.5\,\text{K}$ and a magnetic field of $\mu_0 H=0.15~\text{T}$ along the surface normal; the bottom green line showcases the substantial XMCD signal. In the inset we show that the peak remanent XMCD signal scales inversely with $\theta$, where $\theta$ is the angle between the magnetization direction and the x-ray beam. This decline of XMCD is a strong indication of an out-of-plane easy axis for the Mn moments.  

In Fig.~\ref{fig:XMCD}b we show the magnetization obtained by measuring the peak $L_3$ XMCD signal at $T\approx3.5\,\text{K}$ within a field range of $\pm0.3\,\text{T}$. It exhibits a substantial remanence at $\mu_0 H=0\,\text{T}$, in sharp contrast to MnBi$_2$Te$_{4}$, which exhibits no remanent magnetization, and MnBi$_4$Te$_{7}$, which has a smaller remanence-to-saturation ratio \cite{PhysRevX.9.041065}. Furthermore, we observe a coercive field of $\mu_0 H_c=45\,\text{mT}$. We caution against overinterpreting the similarity of this $H_c$ with the bulk one: First, the data were measured at somewhat different temperatures, which has an effect on $H_c$ (Fig. \ref{fig:SQUID}). Second, different ramping speeds were used, which, too, has an effect on $H_c$ for magnetic TIs \cite{Golias21_APL119_132404}. In addition, the hysteretic behavior of surface and bulk might be intrinsically different.

Finally, in Fig.~\ref{fig:XMCD}c we compare the $T$-dependent remanent peak $L_3$ XMCD signal  measured with surface sensitive TEY with the one measured with bulk sensitive total fluorescence yield (FY). Within the precision allowed by the $T$ increments of $2\,\text{K}$, the transition temperatures at surface and bulk are consistent. We remark that the transition behavior as observed with SQUID and XMCD could differ somewhat due to the different measurement protocols: For XMCD, each point in Fig.~\ref{fig:XMCD}c  was obtained after driving to $\mu_0 H = 3\,\text{T}$ and back to remanence. In contrast, in SQUID measurements a conventional FC protocol at $10\,\text{mT}$ was used.

\subsection{MLFT Calculations}

\label{sec:MLFT}

\begin{table*}
\renewcommand*{\arraystretch}{1.5}
\caption{\label{tab:table1} Slater integrals obtained from DFT and spin--orbit coupling constants in the Hartree-Fock approximation for the Mn$^{2+}$ ion (in units of eV).}
\begin{tabular}{l l c S[table-format=4.3]S[table-format=4.3]S[table-format=4.3]S[table-format=4.3]S[table-format=4.3]S[table-format=4.3]S[table-format=4.3]}
\hline\hline
{ion} &{state } & {\quad configuration\quad } &{\quad $F^{(2)}_{dd}$} & {\quad $F^{(4)}_{dd}$} & {\quad $\zeta_{3d}$} & {\quad $F^{(2)}_{pd}$} & {\quad $G^{(1)}_{pd}$} & {\quad $G^{(3)}_{pd}$} & {\quad $\zeta_{2p}$} \\
\hline
{Mn$^{2+}$\quad }  & {initial} &{\quad2$p^{6}$3$d^5$ } & 9.4323 & 5.8132 & 0.040 &      &       &       & \\
{ }               & {final}   &{\quad2$p^{5}$3$d^6$ } & 10.1963 & 6.2899 & 0.053 & 5.3354& 3.8379 & 2.1773 & 6.846 \\
\hline

\hline\hline
\end{tabular}
\label{table:Slater}
\end{table*}

 The line shapes of the XAS and XMCD spectra contain important physical information, such as the $d$-electron configuration, including the local magnetic moments. Therefore, we have modeled our experimental data by MLFT (multiplet ligand-field theory). In our approach \cite{PhysRevB.85.165113, PhysRevB.96.245131, Tcakaev, Tcakaev20_Eu} (sec. \ref{sec:METHODS}), rather than relying on oversimplified approximations, we adjust most of the MLFT parameters to the data and obtain $10Dq=0.06$, $10DqL = 2 T_{pp}=1.9$, $\Delta=1.1$, $U_{dd}=4.0$, $U_{pd}=5.0$, $V_{e_g}=1.3$ and $V_{t2_g}=0.65$ (all in units of eV). For the SO coupling constants, we use the Hartree-Fock values \cite{Havercortthesis}, whereas the Slater integrals are calculated based on DFT in the local density approximation (LDA, table~\ref{table:Slater}).
 
\begin{figure}[b]
	\centering
\includegraphics[width=0.8\columnwidth]{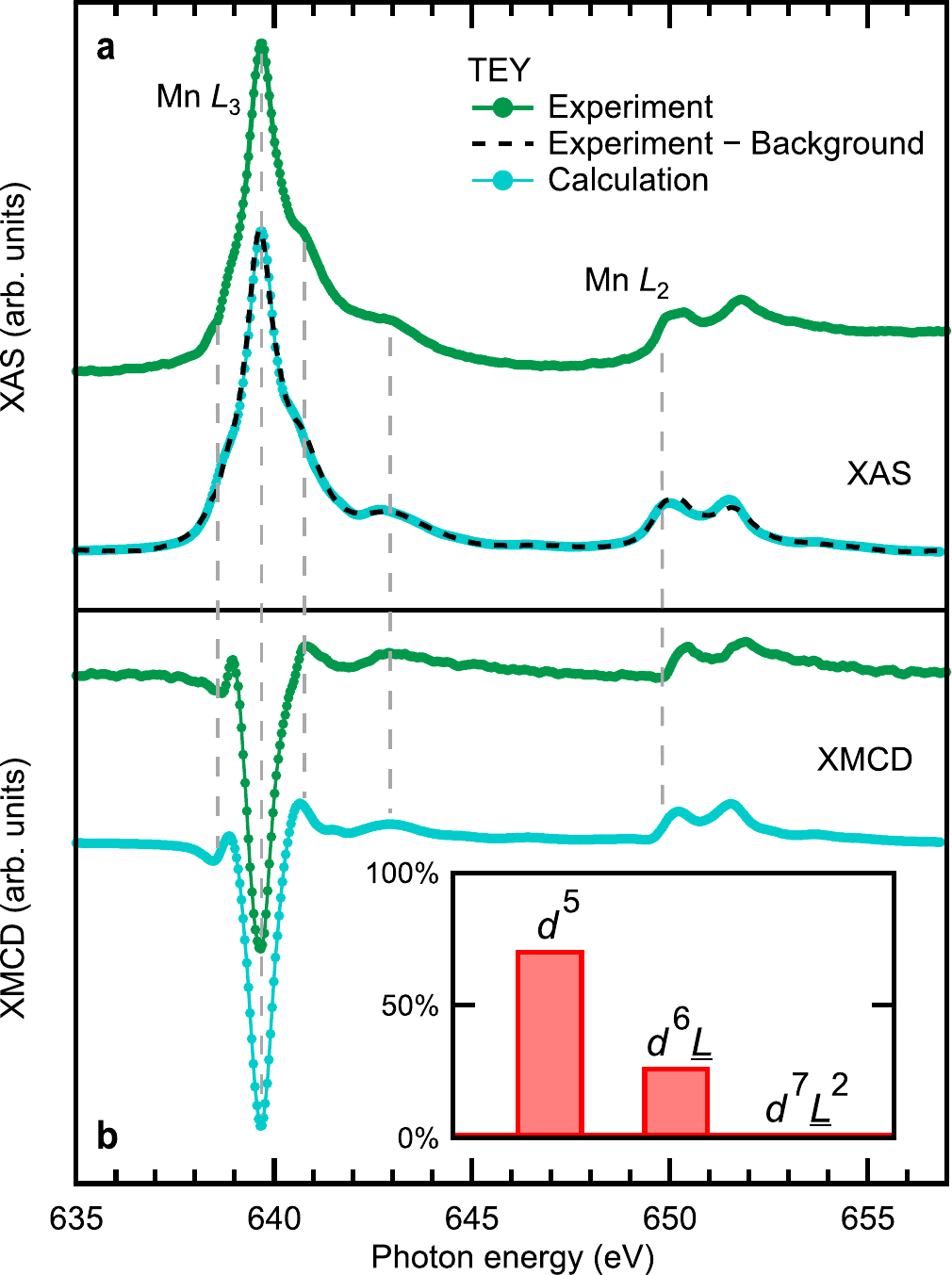}
	\caption{MLFT analysis. (a) Background-corrected polarization-averaged experimental XAS spectrum (dashed line) together with a calculated MLFT spectrum (blue). The original, uncorrected data is shown above (green line). (b) Corresponding experimental and calculated XMCD spectra. The inset shows the contributions of different electronic configurations to the ground state. The vertical dashed lines highlight the positions of particular features of the spectra.
	}
	\label{fig:MLFT}
\end{figure}

The calculated spectra (Fig.~\ref{fig:MLFT}) show an excellent agreement with the experimental data, most notably for the XMCD, reproducing all the multiplet features and their relative energy positions. Whereas the nominal Mn$^{2+}$ $d^5$ configuration ($^6S_{5/2}$) dominates with $71\%$, there is significant charge transfer from the Te ligands, resulting in a $27\%$ contribution of $d^6\underline{L}$ to the ground state ($d^7\underline{L}^2$ contributes negligibly, see the inset in Fig.~\ref{fig:MLFT}b). This hints towards a considerable hybridization between Mn $d$ and ligand $p$ orbitals. The resulting $3d$ electron filling is $n_d=5.31$, corresponding to an effective $1.69+$ valence. We obtain $m_\text{eff}^\text{spin}=\sqrt{\langle \mathbf{m}^2\rangle} = 5.67 \mu_\text{B}$ for the local effective moment, as well as $m^\text{spin}=g_s \langle S_z \rangle =4.68 \mu_\text{B}$ and $m^\text{orb}=g_l\langle L_z \rangle=0.008 \mu_\text{B}$ for the maximal $z$-projections of the spin and orbital moments, respectively.

Finally, we observe that the electron filling and the magnetic moments resulting from our MLFT analysis are in excellent agreement with those calculated based on the DFT-GGA+U calculations in sec.~\ref{sec:DFT-GGA}: We obtain $n_d=5.3$ and $m^\text{spin}=4.7 \mu_\text{B}$ using the same interaction parameters as in MLFT. These results are also in good agreement with the bulk magnetometry data (see sec.~\ref{sec:SQUID} and Fig.~\ref{fig:SQUID}), as well as with published neutron diffraction data \cite{PhysRevMaterials.4.054202, PhysRevB.101.020412}.

\begin{figure}
	\centering
       \includegraphics[width=0.85\columnwidth]{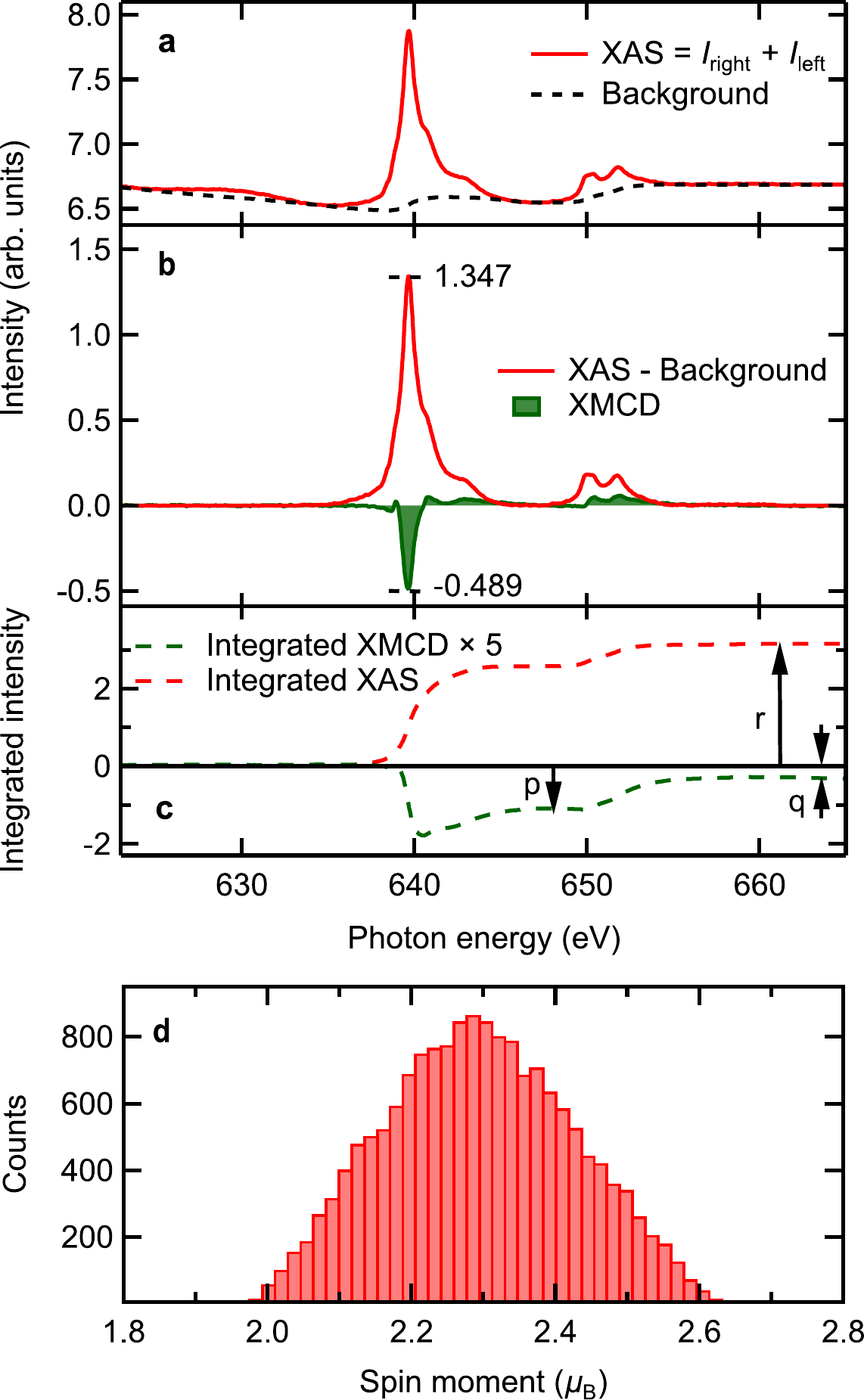}
	\caption{Sum rule and peak asymmetry analysis of data measured on sample $\#4$ at $T\approx3.5\,\text{K}$ and $\mu_0 H = 0.15\,\text{T}$. (a) Polarization-averaged XAS intensity $I$ (red line) together with the background (dashed line). (b) XAS spectrum after background correction, together with the XMCD signal $I_\text{XMCD}$ (filled green curve). The peak intensities necessary for the asymmetry analysis are marked. (c) Integrated intensities of the XAS and the XMCD (multiplied by $5$) spectra. The integrals $p$, $q$ and $r$ necessary for the sum rules are indicated with arrows. (d) Distribution of $m^\text{spin}_\text{XM}$ obtained by randomly varying the sum rule parameters within reasonable error margins, but not considering the uncertainty in the background choice (sec.~\ref{sec:SumRules} and sec.~S.V).
	}	
	\label{fig:analysis}
\end{figure}

\subsection{XMCD Sum Rule and Peak Asymmetry Analysis}

\label{sec:SumRules}

The MLFT analysis yields the \textit{local} magnetic moments based on the spectral \textit{shape}. The sum rules, in turn, relate the integrated  Mn $L_{2,3}$ XMCD and x-ray absorption spectral \textit{intensity} to the \textit{long-range ordered} orbital and spin magnetic moments near the surface \cite{TholeOrbSumRule, TholespinSumRule, PhysRevLett.75.152} (sec.~\ref{sec:METHODS} and sec.~S.V-A). We show their application to data for sample $\#4$ in Fig. \ref{fig:analysis}. After a background correction (Fig. \ref{fig:analysis}a), we obtain the XAS and XMCD data (Fig. \ref{fig:analysis}b), from which we calculate the integrals $p$, $q$ and $r$ required for the sum rule analysis (Fig. \ref{fig:analysis}c). Finally, in Fig. \ref{fig:analysis}d we show the distribution of the resulting values for $m_\text{XM}^\text{spin}$ obtained by applying the analysis $16384$ times while randomly varying the sum rule parameters within reasonable error margins. Also taking into account some ambiguity in the choice of the XAS background due to the rather featureless but intense tails of the preceding Te $M_{4,5}$ edges allows us to estimate the errors (sec.~S.V). We obtain $m_\text{XM}^\text{spin}=2.3\pm0.25$, $m_\text{XM}^\text{orb}=0.1\pm0.15$ and $m_\text{XM}^\text{tot}=2.4\pm0.3$ (in $\mu_\text{B}$/Mn, table \ref{table:param}). The same analysis for sample $\#1$ yields $m_\text{XM}^\text{tot} = (2.2\pm0.35)\mu_\text{B}/\text{Mn}$, which is compatible with sample $\#4$ within the error.

An alternative way to obtain $m_\text{XM}^\text{spin}$ is to analyze the XMCD $L_3$ peak asymmetry \cite{PhysRevB.56.5461,Edmonds_2015,FIGUEROA201793}, which avoids the problems arising from uncertainty in $p$ due to the overlap of the $L_{3}$ and $L_{2}$ peaks (sec. S.V-B). We obtain $m_\text{XM}^\text{spin} = (2.55\pm0.25)\mu_\text{B}/\text{Mn}$, which is about $10\%$ larger than the sum rule result. 

Table \ref{table:param} also shows that the orbital moment is negligible within the error, as expected for a predominant $d^5$ configuration (sec.~\ref{sec:MLFT}). The total moment $m^\text{tot}_\text{XM}$ obtained with surface sensitive XMCD is reduced by about $40\%$ in comparison to the one obtained with bulk sensitive SQUID magnetometry. Increasing the field to $6\,\text{T}$ brings the moment $m_\text{XM}^\text{tot}$ to $3.9\,\mu_\text{B}$, i.\,e. closer to $m_\text{SQ}^\text{tot}=4.2\,\mu_\text{B}$ and to the theoretical maximal moment (sec.~\ref{sec:MLFT}).

It is important to keep in mind that the indicated errors of the XMCD results only take into account statistical fitting and background estimate effects. However, the shallow probing depth can further bias the outcome (sec.~\ref{sec:XASXMCD} and ref. \cite{doi:10.1063/1.4904900}): It is reasonable to expect that the QLs and the outer ($6c$) positions of the SLs might contain slightly canted Mn, as well as a few percent of paramagnetic and possibly even AFM (with respect to the $3a$ positions) Mn, see sec.~\ref{sec:growth}, sec.~\ref{sec:discussion} and Fig. \ref{fig:stack}. Already for SL surface termination, the FM ordered Mn sheet ($3a$ positions) is buried about $0.55\,\text{nm}$ below the surface. At the same time, there is Mn in $6c$ positions closer to the surface, both in the very same SL and in QLs, which can terminate the surface in different parts of the sample. Therefore, already for a mean probing depth (MPD) of about $1\,\text{nm}$, the contribution of the FM Mn would be notably suppressed. MPD values between $1$ and $2.5$ have been reported for this photon energy range \cite{Frazer03_SurfSci03_537_161, Abbate92_SurfInterfAnal18_65}. Given the presence of the heavy elements Te and Bi, which might even further attenuate the escaping electrons, an MPD close to $1\,\text{nm}$ is not unrealistic and the involvement of probing depth effects is well conceivable.

\section{Discussion}
 
\label{sec:discussion}

The major finding of our study is that the surface of MnBi$_6$Te$_{10}$ exhibits FM properties comparable to its bulk, with a robust FM subsystem in the topmost septuple layer, which can interact with the topological surface states. Indeed, recent ARPES reports suggest the opening of an exchange gap of about $15\,\text{meV}$ \cite{yan2021delicate}. 

\begin{table}
\renewcommand*{\arraystretch}{1.5}
\setlength\tabcolsep{1.6pt}
\caption{\label{tab:SUmRuleSumm} Comparison of the magnetic moments obtained from analysis of surface sensitive XAS and XMCD data measured at $T\approx 3.5$ K in a $0.15\,\text{T}$ field with bulk-sensitive SQUID magnetometry results at $T=2$ K and the same field (in units of $\mu_\text{B}$/Mn). The error bars are explained in Supporting Information sec V.}
	\label{table:param}
	\begin{ruledtabular}
		\begin{tabular}{lclc}
	Sum rules:			& $m^\text{spin}_\text{XM}$     &	$2.3\pm 0.25$	 \\
						& $m^\text{orb}_\text{XM}$	  &	$0.1\pm0.15$   \\ 
						& $m^\text{tot}_\text{XM}$	  &	$2.4\pm0.30$  \\ \hline	
															
	Asymmetry:			& $m^\text{spin}_\text{XM}$	  &	$2.55\pm0.25$ \\ \hline	
	
	SQUID:				& $m^\text{tot}_\text{SQ}$ 	  &	$4.2\pm0.2$	 \\	 
		\end{tabular}
	\end{ruledtabular}
\end{table}

As outlined in sec.~\ref{sec:SumRules}, probing-depth effects can at least partially explain the $40\%$ reduction of the XMCD-derived remanent moment as compared to the SQUID-derived bulk moment.  Additional surface effects might influence magnetism and therefore warrant consideration. First of all, the  incomplete out-of-plane coordination by magnetic neighbors of the topmost SL suppresses the out-of-plane magnetic interactions and interrupts the exchange paths of the antisite Mn ions. However, since these interlayer interactions are weak (sec.~\ref{sec:DFT-GGA}), additional theoretical scrutiny would be required to elucidate, what role their further suppression might play. Second, a competition between the demagnetizing field and the crystalline anisotropy might result in a canting and a suppression of the XMCD signal -- an effect which must, however, be small due to our finding of a strong out-of-plane anisotropy, see sec.~\ref{sec:XASXMCD} and the inset of Fig.~\ref{fig:XMCD}a. Third, a combined study involving DFT and XMCD  suggests that TSS couple to magnetic atoms such as Co and Mn at the surface of Bi$_2$Te$_3$, contributing to their interaction by a RKKY-like mechanism \cite{Ruessmann18_JPhysMater1_015002}: Due to their highly localized nature, electrons in the TSS interact more strongly with magnetic moments than electrons in the bulk. This might contribute to the differences between the magnetic properties at the surface and in the bulk. Our results encourage similar calculations for MnBi$_6$Te$_{10}$. Finally, the interaction with the TSS might also result in a slight helical canting away from the out-of-plane orientation, driven by Dzyaloshinskii--Moriya interactions \cite{PhysRevResearch.3.033173,PhysRevResearch.3.L032014}. Again, due to the strong out-of-plane anisotropy we experimentally observe, this contribution would be small.

We now discuss the mechanism inducing the crossover from a pronounced AFM towards an FM order as the number $n$ of the QLs in the MBT$_n$ stacking sequence increases (Sec.~\ref{sec:SQUID} and Fig.~\ref{fig:SQUID}). Our calculations for the ordered MnBi$_6$Te$_{10}$ (Sec.~\ref{sec:DFT-GGA}) yield an -- albeit small -- AFM coupling. Although the increasing $K/2J$ ratio with increasing $n$ \cite{PhysRevLett.124.197201} certainly helps to stabilize the FM order for $n\geq n_\text{FM}=2$ in our samples (see Sec.~\ref{sec:SQUID} and Fig.~\ref{fig:SQUID}c), the fact that previous studies reported $n_\text{FM}=3$ \cite{Hueaba4275, Klimovskikh2020,Xie2020,PhysRevB.102.035144} hints at an additional phenomenon being involved. In sections \ref{sec:growth} and \ref{sec:DFT-GGA} we have established the presence and the role of Mn/Bi antisite defects that can drive enhanced FM properties. As our numerical modelling has shown both for the MBT$_0$ with the strongest interlayer AFM coupling and for the MBT$_1$ with a periodic alternation of SLs and QLs, the motif of an intermixing pattern determines whether ferro- or antiferromagnetism is preferred. Hence, the observed magnetic properties of our MnBi$_6$Te$_{10}$ samples likely originate in a prevalence of intermixing patterns that favor the FM order.

In this context, a comparison between the MBT$_n$ series and the analogous Sb-based family (MnSb$_2$Te$_4$)(Sb$_2$Te$_3$)$_n$ (MST$_n$) becomes relevant. The FM order is more dominant even for $n=0$ in MST$_n$ and MBST$_n$, and, as widely  accepted by now, is driven by the Mn/Sb intermixing \cite{wimmer2020ferromagnetic,PhysRevX.11.021033,PhysRevB.103.134403, Xie_2022}. This phenomenon is much stronger in MST$_n$ than in MBT$_n$ since it is facilitated by closer atomic radii of Mn and Sb. On the other hand, the impact of the intermixing-induced FM state in MnSb$_2$Te$_4$ on its band topology is still under ongoing debate~\cite{PhysRevX.11.021033, wimmer2020ferromagnetic, JMat2020, PhysRevLett.122.206401, Natcomm2019, PhysRevB.102.085114, PhysRevB.100.195103, physchem22}. Intrinsic p-type doping in MST$_0$ hampers clear-cut spectroscopic observations of the possible surface states and, thus, an ultimate conclusion about its topological nature. Furthermore, QAHE realizations in the MST$_n$ have not been reported. In fact, Ref.~\cite{PhysRevX.11.021033} argues the importance of further studies on how intermixing impacts bulk and surface magnetism in the established topological MBT$_n$ materials, but focuses on the MST$_0$ instead, because the necessary intermixings were not accessible  by the bismuth analog at that time. So far intermixing in the MBT$_n$ has been discussed mostly in the terms of its influence on the Dirac-point gap~\cite{Garnica2022}, while the consequences for the magnetism are quite unclear. En route to understanding the broader role of intermixing, a recent study reveals its crucial influence on the magnetic coupling in MnBi$_2$Te$_4$~\cite{PhysRevB.103.184429.2021}, and our current work pinpoints the particular antisite defects that enhance (or suppress) the local FM coupling in the MBT$_n$ series.

 We have established that the FM properties of our crystals are conditioned by the underlying cation intermixing. It is instructive to examine whether this relationship holds true for the other published works. Whereas Mn deficiency in MnBi$_6$Te$_{10}$ is often found by x-ray spectroscopy~\cite{Hueaba4275, Xie2020, PhysRevB.102.035144}, the related intermixing has been scrutinized only in~Ref.~\onlinecite{Klimovskikh2020}. On the one hand, their and our samples have such commonalities as the presence of  Mn/Bi intermixing, the absence of cation vacancies, and a strongly mixed occupancy on the $3a$ site. On the other hand, there are also substantial differences: The mixed $3a$ occupancy is more pronounced in our sample, in which we find $56\%$ Mn (and $44\%$ Bi), than in the sample studied in Ref.~\onlinecite{Klimovskikh2020}, which has $83\%$ Mn (and 17~\% Bi). Most importantly, the Mn distribution over the 6\textit{c} positions is distinctly different: We observe a higher Mn concentration in both $6c$ sites of the QLs, i.e. up to Bi$_{1.86}$Mn$_{0.14}$Te$_3$ vs.\ Bi$_{1.92}$Mn$_{0.08}$Te$_3$ in Ref.~\onlinecite{Klimovskikh2020}, and up to 2~\% Mn in the outer positions of the SL that are reported  defect-free in Ref.~\onlinecite{Klimovskikh2020}. In general, the $3a$ site in our crystals is more Mn-depleted, so that these ``stray'' Mn atoms, which find no space on the $3a$ site, disperse over the entire layered stack by occupying $6c$ sites. In  accordance with our theoretical deliberations in Sec.~\ref{sec:DFT-GGA} (models S$_1$ to S$_4$), the less pronounced intermixing and the presence of swapped Mn only in one of the two  $6c$ sites of the QLs in the samples of Ref. \cite{Klimovskikh2020} accords with them featuring an AFM ground state.
 
The question of why intermixing takes place and which kind of defects are more likely to occur is evidently very relevant and, at the same time, a complex one.
First, recent literature has shown that antisite cationic defects have the lowest formation energy and are energetically favorable to form in both MnBi$_2$Te$_4$ and MnBi$_4$Te$_7$~\cite{defects}.
It has been argued that such defects provide an effective way to release a lattice strain effect which occurs within the septuple layer of MnBi$_2$Te$_4$ due to a mismatch between the MnTe and Bi$_2$Te$_3$ structure fragments. 
Second, as argued below, variations in the intermixing patterns of MnBi$_6$Te$_{10}$  samples produced by different groups may stem from subtle differences in the synthetic procedures, pointing to the relevance of finite temperature effects for the relative stability of different defects. This would be no surprise since these compounds are formed at elevated temperatures and are metastable at room temperature as we have previously shown \cite{C9TC00979E,Zeugner2019}.

Comparing our growth conditions (see Sec. \ref{sec:METHODS}) for MnBi$_6$Te$_{10}$ to those of Refs.~\onlinecite{PhysRevB.100.155144, Klimovskikh2020, PhysRevMaterials.4.054202} reveal differences in dwelling times, the starting and  quenching temperatures, and the composition of a melt, which may well account for the various intermixing patterns. 
In general, we observe reproducible Mn concentrations and magnetic behavior in our MBT$_2$ crystals (Fig.~\ref{fig:SQUID} and Fig.~S4) for an applied tempering profile~\cite{folkers2021, C9TC00979E}, suggesting that the cation intermixing is a temperature-regulated phenomenon. We do not argue that this process is fully governed by the thermodynamics, since crystallization of the MBT$_n$ from a heterogeneous melt is strongly kinetics-driven. Yet it seems plausible, that the resulting intermixing pattern is governed by a given synthetic protocol. Strong correlations between the synthesis temperatures and the resultant cation disorder and magnetic order have been, by now, undoubtedly established at least for MnSb$_2$Te$_4$~\cite{wimmer2020ferromagnetic, Liu2020, folkers2021}.  Since the Bi-analogs have limited thermodynamic stability~\cite{Zeugner2019, C9TC00979E} and, thus, offer very narrow growth temperature windows, their degrees of intermixing appear to be far less dramatic than in MnSb$_2$Te$_4$ and, therefore, more challenging to trace experimentally. Less substitutional disorder than in the MST$_n$ (on average) may be a blessing when it comes to optimizing an MBT$_n$ material's system for the QAHE device fabrication~\cite{Deng2021}.

In summary, the prominent ferromagnetic characteristics of our sample, with a rather large $T_c$, and a substantial ordered, out-of-plane moment both in the bulk and at the surface, categorizes MnBi$_6$Te$_{10}$ as a particularly interesting candidate for the realization of a high-temperature QAH material~\cite{Deng895, Deng2021, Hu2020, PhysRevB.102.035144}. Moreover, a monolayer of ferromagnetic MnBi$_6$Te$_{10}$ appears as a perspective candidate for magnetic extension~\cite{ferroext,Otrokov_2017} and proximity setups, since an FM MnBi$_6$Te$_{10}$ slab was predicted to exhibit QAHE~\cite{PhysRevLett.123.096401}.

\section{METHODS}
\label{sec:METHODS}

\subsection{Crystal growth and characterization}

Pre-synthesized, phase-pure MnTe and Bi$_2$Te$_3$ powders were mixed in a ratio $0.85:2$ at.~$\%$, pelletized and placed in an evacuated quartz tube. This was inserted at $T=923\,\text{K}$ into a preheated two-zone tube furnace with temperature control via external thermocouples (Reetz GmbH). The ampule was subsequently cooled down to 858 K at a rate of 1 K/hour, tempered for 14 days and then quenched in water. Platelet-like MnBi$_6$Te$_{10}$ crystals (lateral size up to 1 mm) were mechanically separated from the obtained ingot.

 Powder x-ray diffraction data were collected on an X'Pert Pro diffractometer (PANalytical) with Bragg-Brentano geometry (featuring variable divergence slits) operating with a curved Ge(111) monochromator and Cu-K$\alpha_1$ radiation ($\lambda = 154.056\,\text{pm}$). The phase composition of the polycrystalline ingot and individual crystals was estimated by Le Bail or Rietveld methods in JANA2006 \cite{Petricek}. The preferred orientation of the crystallites was described by March-Dollase corrections, the roughness for the Bragg--Brentano geometry was accounted for by the Suorti method.

Scanning electron microscopy (SEM) was performed using a SU8020 (Hitachi) equipped with a X-MaxN (Oxford) Silicon Drift Detector (SDD) at $U_a = 2-5\,\text{kV}$. The composition of selected single crystals was determined by semi-quantitative energy dispersive x-ray analysis at $20\,\text{kV}$ acceleration voltage. 

\subsection{Bulk magnetometry measurements} 

Field and temperature dependent magnetization studies were performed using a Quantum Design superconducting quantum interference device (SQUID) magnetometer equipped with a vibrating sample magnetometer (VSM) option (MPMS3). Our magnetization data on samples $\#1 -\#4$ are normalized to the real compositions determined via EDX. To obtain the absolute magnetization $M$ per Mn atom, a precise knowledge of the sample mass is important. Samples $\#1$ and $\#4$ have an approximately 10 times smaller mass than sample $\#2$, increasing the error of $M$. Nevertheless, the data for all four samples agree well with each other (Fig.~S4). Furthermore, we here refer to the projection of the total magnetic moment onto the $z$-axis ($\mathbf{H}||\mathbf{z}$), which is the quantity obtained from the SQUID magnetometry measurements, as $m_\text{SQ}^\text{tot}$.

A setup made of two half-cylindrical quartz rods fixed with a small quantity of GE varnish to the main quartz VSM sample holder was designed to ensure an alignment of the crystals such that the external magnetic field was applied perpendicular to the crystal surface. Note that this setup, however, results in a rather temperature-independent (at not too low temperature) but non-negligible background contribution to the magnetic susceptibility, hindering a reliable extraction of the Curie-Weiss constant $\theta_{\textrm{CW}}$ and the temperature independent susceptibility $\chi_0$ for our low-mass samples MnBi$_6$Te$_{10}$.

\subsection{Bulk DFT (GGA$+U$) calculations}

Fully relativistic DFT calculations based on the Generalized Gradient Approximation + $U$ were performed with the parametrization of Perdew, Burke, and Ernzerf \cite{perdew1996generalized}, using the full localized limit for the double-counting correction with $U=U_{dd}$ (the latter as obtained in Sec. \ref{sec:MLFT}) and $J=\big(F^{(2)}_{dd} + F^{(4)}_{dd}\big)/14$, with $F^{(2)}_{dd}$ and $F^{(4)}_{dd}$ the Slater integrals for the initial states presented in Table \ref{table:Slater}. The spin--orbit coupling was included in the four-component formalism as implemented in FPLO. The total energy difference between the FM and A-type AFM configurations was computed using for Brillouin zone integrations a linear tetrahedron method. For MnBi$_6$Te$_{10}$, we use a mesh of the Brillouin zone having $14\times14\times14$ subdivisions. The magnetic anisotropy energy was calculated in the AFM state based on a mesh having $10\times10\times10$ subdivisions. The Mn $3d$ occupancy and the spin projection presented in the main text correspond to the gross projections. For the calculations of MnBi$_4$Te$_7$ based on a $2\times1\times 2$ supercell, we use a mesh with $6\times12\times2$ subdivisions.

\subsection{XAS and XMCD measurements}

The XAS and XMCD measurements were performed using the high-field cryomagnet end station HECTOR of the BOREAS beamline at the ALBA synchrotron radiation facility \cite{ALBABOREAS} and at the high-field diffractometer at the UE46 PGM-1 beamline, BESSY II \cite{BESSY}. The single crystals were glued with conducting silver epoxy onto Cu sample plates and mounted on the cold finger of a helium flow cryostat. Prior to the measurements, each sample was mechanically cleaved in the fast-entry chamber at a pressure of $\sim$10$^{-9}$~mbar to expose a pristine surface. The sample was then transferred into the spectroscopy chamber with a pressure in the 10$^{-11}$ --  10$^{-10}$ mbar range. 

The measurements were carried out in the TEY or FY mode at magnetic fields of up to $6\,\text{T}$ and at various temperatures in the $3.5\text{--}35~\text{K}$ range. The temperature was calibrated with a thermal sensor mounted at the sample position before the experiment. Especially below about $5\,\text{K}$, the actual sample temperature crucially depends on the thermal contact, increasing its error as compared to higher temperatures. The spectral intensity was normalized by the incoming photon intensity ($I_0$). We used circularly polarized light at both beamlines. The area probed by the beam at both  facilities (about $120\times80 \mu \text{m}^2$) was much smaller than the sample size. 

The raw XAS spectra were scaled with respect to each other to have the same intensity at energies far from the resonances to obtain $I_\text{left}$ and $I_\text{right}$. We define the XMCD signal as $I_\text{XMCD} = I_\text{left} - I_\text{right}$. The average, not background corrected XAS is $I = (I_\text{left}+I_\text{right})/2$. To cancel out any experimental drifts, for each data set we measured eight spectra in a row by altering the X-ray polarization. Finally, the magnetic moments measured with XMCD are marked with the subscript XM, e.\,g. $m_\text{XM}^\text{tot}$.

\subsection{MLFT calculations}

As a starting point to obtain input parameters for the MLFT modeling, self-consistent DFT in the linear density approximation is sufficient, which we performed using the FPLO package \cite{PhysRevB.59.1743}. The Brillouin zone was sampled by a $2\times2\times2$ $k$-point mesh. The exchange-correlation potential was treated in LDA, with the scalar relativistic functional according to Ref. \onlinecite{PhysRevB.45.13244}. We have used the experimental crystal structure from Ref. \onlinecite{C9TC00979E}: rhombohedral space group $R\bar{3}m$ (166), $a=4.37$ {\AA} and $c=101.83$ {\AA}, slightly distorted octahedral Mn coordination with Mn-Te bond length of $3.00$~{\AA} ($C_{3v}$ crystal field symmetry). We obtained Wannier orbitals as input for MLFT by downfolding to a basis set of Mn $3d$, Te $5p$ and Bi $6p$ orbitals in an energy window from $-6$ to 3 eV including an exponential decaying tail with a decay of 1 eV at the boundaries of the selected energy range.

The MLFT calculations  were performed using the Quanty package \cite{PhysRevB.90.085102, PhysRevB.85.165113, Haverkort_2014} within the CI scheme considering the nominal $2p^63d^5$ (Mn$^{2+}$) configuration and two further charge-transfer states $d^6\underline{L}$ and $d^7\underline{L}^2$. The spectral contributions from the split ground-state terms were weighted by a Boltzmann factor for $T = 2\,\text{K}$. The mean-field effective potential was modeled by an exchange field estimated from the $T_\text{c}$ of $12\,\text{K}$. Instrumental and lifetime effects were taken into account by a Gaussian broadening of $0.35\,\text{eV}$ (FWHM) and an $E$-dependent Lorentzian profile of $0.15-0.35\,\text{eV}$ (FWHM).

The Slater integrals for the MLFT calculations were obtained by DFT, where $F^{(2)}_{dd}$ and $F^{(4)}_{dd}$ were scaled up by $8\%$ for the final state, improving the agreement to experiment. SO coupling constants were kept to the Hartree-Fock values \cite{Havercortthesis}. $\Delta=E(d^{n+1}\underline{L})-E(d^n)$, $U_{dd}$ and $U_{pd}$ were directly fitted to the experimental spectra, keeping $U_{dd}$/$U_{pd}=0.8$ \cite{PhysRevB.53.1161, PhysRevB.46.3771, PhysRevB.45.1561, PhysRevB.30.957}. Experiments involving charge-neutral excitations such as XAS are only weakly sensitive to $\Delta$, $U_{dd}$ and $U_{pd}$. In our particular case we were fitting simultaneously XAS and XMCD spectra, which substantially mitigates these kind of problems. Our results are in good agreement with values reported for (Ga,Mn)As \cite{PhysRevLett.107.187203, PhysRevLett.107.197601,PhysRevB.59.R2486, doi:10.1063/1.1751619, Pincelli2017} and Mn-doped Bi$_2$Se$_3$ \cite{doi:10.1063/1.4904900} and Bi$_2$Te$_3$ \cite{Watson_2013}. The other MLFT input parameters were estimated from DFT, and their values were subsequently adjusted to reproduce the experimental spectra. To simplify the calculation, instead of the trigonal $C_{3v}$ we work in $O_h$ symmetry, with the $C_4$ octahedral axes along the Mn-Te bonds, which has a negligible impact: Our simplification neglects the splitting of the $t_{2g}$ orbitals, which is tiny compared to $10Dq<100\,\text{meV}$, which in turn is smaller than the experimental resolution.

\subsection{XMCD Sum Rule and Peak Asymmetry Analysis}

The XMCD sum rules yield:

\begin{equation}
m^\text{orb}_\text{XM}=-\frac{4}{3}\frac{q}{r}(10-n_d),
\end{equation}

\begin{equation}
m^\text{spin}_\text{XM}=-\frac{6p-4q}{r} (10-n_d) C +7\langle T_z \rangle,
\end{equation}

where $p$ and $q$ are the XAS intensity differences $(I_\text{left}-I_\text{right})$ integrated over the $L_3$ edge and the entire $L_{2,3}$ region, respectively (Fig. \ref{fig:analysis}). The XAS intensity $I$, after background correction (sec.~S.V-A), is integrated over $L_{2,3}$ to yield $r$. $\langle T_z \rangle$ is the expectation value of the intra-atomic magnetic dipole operator which is $-0.0002\hbar$ and hence negligible (sec.~S.V-A). For $n_d$ we use the MLFT value of 5.31 (sec. \ref{sec:MLFT}). Finally, $C$ is a correction factor, which takes into account the considerable overlap of the $L_3$ and $L_2$ contributions for light transition metals. We use a value of $C=1.4$ (sec.~S.V-A). To circumvent the difficulties related to this overlap, one can obtain $m^\text{spin}_\text{XM}$ by a comparison of the experimental XMCD asymmetry at the $L_3$ peak to the theoretical one calculated from MLFT spectra of comparable line width \cite{PhysRevB.56.5461,Edmonds_2015,FIGUEROA201793}.
\newline

\section{Authors contributions}

L.F., E.K. and A.I. conducted and analyzed the experimental work related to crystal growth, XRD and EDX. B.R., L.T.C. and A.U.B.W. performed and analyzed bulk magnetometry measurements. J.I.F. and J.v.d.B. performed and analyzed density-functional calculations aimed to understand the effects of intermixing. A.T., V.B.Z., T.R.F.P., P.G., M.V. and E.W. performed the XAS measurements. A.T. and V.B.Z. analyzed the XAS data with input and support from T.R.F.P., P.K., S.H., H.B., R.J.G., M.W.H., and V.H. V.B.Z., A.I. and V.H. conceived and supervised the project with input and support from F.R., J.v.d.B., B.B. and A.U.B.W. All authors contributed to the discussion and writing of the manuscript.

\section{Acknowledgments}

This work was supported by the Deutsche Forschungsgemeinschaft DFG, Project No.  258499086, SFB 1170 (projects C06 and A01) and Project No.  247310070, SFB 1143. Furthermore, we acknowledge support by the  Deutsche Forschungsgemeinschaft DFG  through the W\"urzburg-Dresden Cluster of Excellence on Complexity and Topology in Quantum Matter -- \textit{ct.qmat} (EXC 2147, project-id 390858490). J.I.F. acknowledges the support from the Alexander von Humboldt Foundation. L.T.C. is funded by the DFG (project-id 456950766). MV and PG acknowledge additional funding by grants PID2020-116181RB-C32, FlagEra SOgraphMEM PCI2019-111908-2 (AEI/FEDER, UE). We acknowledge provision of beamtime at the ALBA synchrotron via proposals 2019093897, 2020024350. We acknowledge financial support and the provision of beamtime by the Helmholtz-Zentrum Berlin. We acknowledge experimental support by Manaswini Sahoo for the SQUID magnetometry studies on MnBi$_8$Te$_{13}$.

\bibliographystyle{apsrev4-1}

%

\end{document}